\documentclass[11pt]{article}

\usepackage{geometry}
\usepackage{graphicx}
\usepackage{amssymb,amsfonts,amsmath}
\usepackage{epstopdf}
\usepackage[footnotesize]{caption}
\usepackage{capt-of}

\usepackage{booktabs}
\usepackage{amsmath}
\usepackage{color}
\usepackage[normalem]{ulem} 

\newcommand{\Kid}{K_\mathrm{id}}
\newcommand{\ploop}{p_\mathrm{loop}}

\newcommand{\J}{J_\mathrm{loop}}

\newcommand{\Rmax}{[R]_\text{max}}

\begin{document}

\title{Sequence Dependence of Transcription Factor-Mediated DNA Looping}

%

\author{%
Stephanie Johnson,\\Department of Biochemistry and
    Molecular Biophysics, \\California Institute of
    Technology, \\1200 E. California Blvd, Pasadena, CA 91125
\and Martin Lind\'{e}n,\\Department of Physics, \\California Institute of
    Technology, \\1200 E. California Blvd, Pasadena, CA 91125\\
    Present address: Center for Biomembrane Research,\\
      Department of Biochemistry and Biophysics, \\Stockholm University,
      Stockholm, Sweden. 
\and Rob Phillips, \\Departments of Physics and Bioengineering, \\California Institute of
    Technology,\\ 1200 E. California Blvd, Pasadena, CA 91125
\footnote{To whom correspondence should be addressed.
Tel: +1 626 395 3374; Fax: +1 626 395 5867; Email: phillips@pboc.caltech.edu}}
\date{}


\maketitle

\begin{abstract}

DNA is subject to large deformations in a wide range of biological processes.  Two key examples  illustrate how such deformations influence the readout of the genetic information: the sequestering of eukaryotic genes by nucleosomes, and DNA looping in transcriptional regulation in both prokaryotes and eukaryotes.  These kinds of regulatory problems are now becoming amenable to systematic quantitative dissection with a powerful dialogue between theory and experiment.  Here we use a single-molecule experiment in conjunction with a statistical mechanical model to test quantitative predictions for  the behavior of  DNA looping at short length scales, and to determine how DNA sequence affects looping at these lengths.  We calculate and measure how such looping depends upon four key biological parameters: the strength of the transcription factor binding sites, the concentration of the transcription factor, and the length and sequence of the DNA loop.  Our studies lead to the surprising insight that sequences that are thought to be especially favorable for nucleosome formation because of high flexibility lead to no systematically detectable effect of sequence on looping, and begin to provide a picture of the distinctions between the short length scale mechanics of nucleosome formation and looping. 

\end{abstract}


\newpage
\section{Introduction}

In its role as the chief informational molecule of the
living world, DNA is subjected to a wide variety of physical
manipulations.  Examples include the looping events that occur during
DNA replication \cite{Echols1990,Schleif1992}, bending of DNA during recombination \cite{Echols1990,Schleif1992}, the bending and twisting induced by a variety of different architectural proteins such
as IHF, H-NS and HU in bacteria \cite{Luijsterburg2006}, the bending induced by the histones
 responsible for packing the genetic material in eukaryotes \cite{Matthews1992,Luger1997}, and the
physical rearrangements of genomic DNA induced by transcription
factors \cite{Echols1990,Schleif1992,Matthews1992,Garcia2007}.  
In fact one of the
most ubiquitous classes of regulatory architecture found in all
domains of life is often referred to as ``biological action at a
distance,'' where transcription factors bind several sites on the DNA
simultaneously, thus looping the intervening DNA \cite{Ptashne1986,Rippe1995,Tolhuis2002}.

Interestingly, many of the biological manipulations experienced by DNA, but especially many cases of ``action at a distance'' in transcriptional regulation,
involve bending and twisting the DNA on length scales that are short in
comparison with its natural scale of deformation, that is, the
persistence length \cite{Garcia2007}.  Eukaryotic DNA is subjected to
enormous deformations when packed in nucleosomes, with 147~bp of DNA
(already smaller than the persistence length) wrapped 1~3/4 times
around the histone octamer \cite{Matthews1992,Luger1997}.  Similarly, in the context of
prokaryotic transcription factor-mediated DNA looping, not only are such lengths
the default in naturally occurring transcriptional networks, but the
optimal {\it in vivo} lengths as determined by the maximal regulatory
effect are often at loop lengths smaller than 100~bp
\cite{Garcia2007,Muller1996}.  
Despite the clear importance of the short length
scale mechanical properties of DNA, however, there remains both
uncertainty and controversy about the ease with which such short DNAs
can be deformed, and also about the role of sequence at these short scales, particularly in the context of protein-mediated bending
(reviewed recently in \cite{Peters2010,Olson2011}).  

\begin{figure}[htbp]
\begin{center}
\includegraphics[width=3.4in]{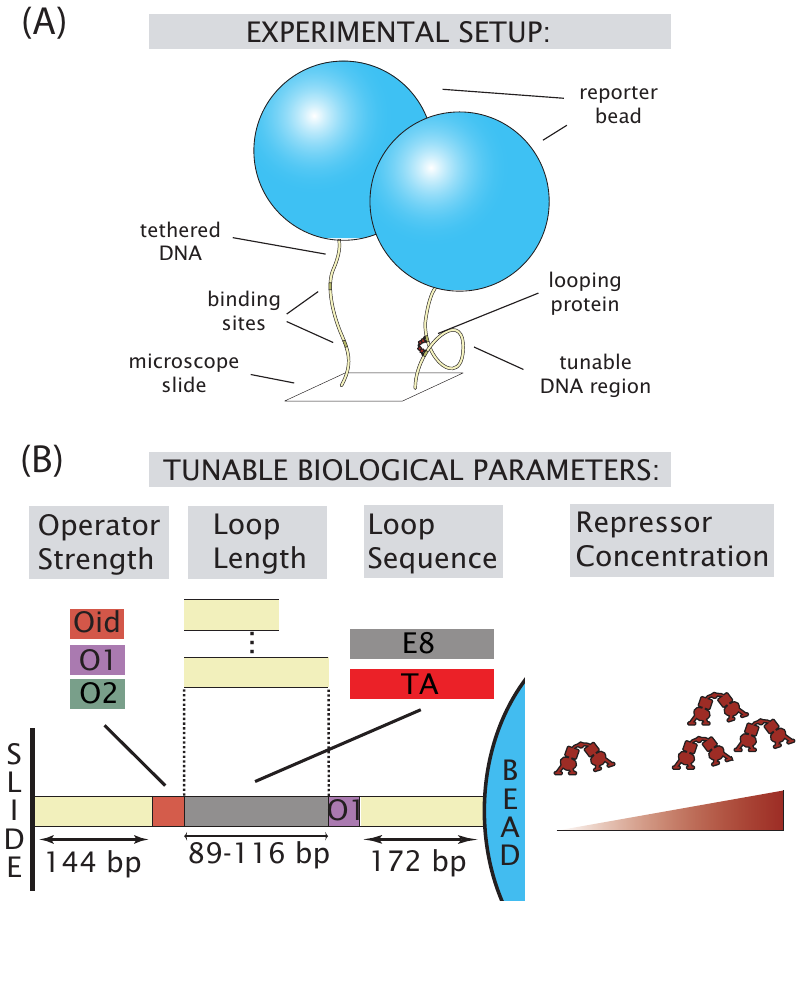}
\end{center}
\caption{Schematic of the Tethered Particle Motion (TPM) assay. {\bf (A)}
  DNA looping is observed as a result of changes in the Brownian
  motion of the tethered bead \protect{\cite{Schafer1991,Yin1994,Finzi1995}}: 
  looping decreases the effective length of the DNA tether, which decreases the bead's root-mean-squared (RMS) motion.  {\bf (B)} Four distinct tunable biological
  parameters varied in the measurements: 1. Repressor binding site, or
  operator.  In this study we use the strong, synthetic ``Oideal''
  ($O_{id}$) operator, the strongest naturally occurring $O_1$ operator,
  and the weaker naturally-occurring $O_2$ operator.  2. Loop
  length. The wild-type {\it lac} operon contains three operators, $O_1$, $O_2$, and an even weaker operator $O_3$, which have the potential to generate three loops of different lengths: the 380 bp $O_1$-$O_2$ loop, the 71 bp $O_1$-$O_3$ loop, 
  and the 472 bp $O_2$-$O_3$ loop.  In our synthetic  constructs we use two operators and 
  systematically tune the distance between them as shown in the figure.   3. Loop sequence. ``E8'' refers to a synthetic random
  sequence, ``TA'' to a synthetic nucleosome positioning sequence (part of the 601TA sequence \protect{\cite{Lowary1998}}). The TA sequence has a higher cyclization J-factor than E8 and is wrapped into nucleosomes {\it in vitro} more readily than E8 \protect{\cite{Cloutier2004,Cloutier2005}}.  4. Lac repressor concentration.  
  }
   \label{fig:ExptSchematic}
\end{figure}

Here we exploit insights about DNA flexibility garnered from one class of genetic regulation where it has been studied extensively, that of nucleosome formation, to make predictions about how a different class of mechanical deformations in regulatory biology, that of DNA looping by a transcription factor, will be altered by these same sequences.  We test these predictions experimentally with a single-molecule assay in conjunction with ideas from statistical mechanics 
for the case of one of the most well known transcriptional regulators in bacteria, that of the Lac repressor, though there are clear implications for other prokaryotic and eukaryotic regulatory motifs as well. 

 As shown schematically in Fig.~\ref{fig:ExptSchematic}, we have combined tethered particle motion (TPM), in which the Brownian motion of a reporter bead is the readout of the state of its
DNA ``leash'' \cite{Schafer1991,Yin1994}, 
with a statistical mechanical model and the systematic variation of four biologically relevant parameters.  The most important of these parameters for the purpose of this study is the flexibility of the DNA in the loop, which is captured in a parameter called the looping J-factor.  The looping J-factor is analogous to the cyclization J-factor obtained in the ligation-mediated cyclization assays that are commonly used to measure DNA flexibility at short lengths, and can be thought of as the effective concentration of one end of the loop in the vicinity of the other \cite{Jacobson1950,Shimada1984}, providing a measure of the energetics of bending the DNA into the loop.  The approach we have developed here allows us to measure these looping J-factors in a way that provides quantitative insights  into how each of the four biologically important parameters we tested affects DNA looping and permits us to contrast the
role of sequence in
DNA cyclization and nucleosome formation with that of looping. We find that two sequences with significantly different propensities for forming DNA minicircles in {\it in vitro} cyclization assays or for forming nucleosomes create a more complicated sequence dependence in the context of DNA loop formation than has been previously appreciated.

\section{MATERIALS AND METHODS}

\subsection{A statistical mechanical model of looping.}

A key tool for making the measurements
presented here is the concentration titration (see Fig.~\ref{fig:E8TA}): by tuning the repressor concentration and measuring the looping probability, we
can fit for other parameters that affect looping probability, namely
the operator dissociation constants ($K_d$'s) and more importantly the looping
J-factors for different DNA sequences and lengths.  Intuitively, at low protein concentrations, the
probability of forming a loop is small.  Similarly, at high
concentrations, the looping probability is low, because the two
operators are each occupied by separate transcription factors.  At
intermediate concentrations, the looping state has its highest
probability.  These intuitions can be captured mathematically
by statistical mechanical models that take into
account all of the different ways that the operators can be decorated
with repressors.  These models make very
strict predictions about the functional form of the looping
probability curves as a function of the various biological parameters.

Our model states that if the
operators have dissociation constants $K_i$ and
$K_{ii}$, and the intervening DNA has looping
J-factor $J_\textrm{loop}$, the looping probability $p_\textrm{loop}$
will be
\begin{equation}
p_\textrm{loop}([R]) = \frac{\frac{1}{2}\frac{[R]J_\textrm{loop}}{K_iK_{ii}}}{
  1+\frac{[R]}{K_i}+\frac{[R]}{K_{ii}}+\frac{[R]^2}{K_iK_{ii}}
  +\frac{1}{2}\frac{[R]J_\textrm{loop}}{K_iK_{ii}}},\label{eqn:model1}
\end{equation}
where $[R]$ is Lac repressor concentration. (See \cite{Han2009} and 
Section~\ref{sec:SImodels} in the Supplementary Material for derivation and 
details.)  
 Although this model was first derived in our earlier work in \cite{Han2009}, as a result
 of the fact that we here explore the analytic consequences of this model, we consider the results presented here to be the first rigorous and successful test of its applicability to DNA looping experiments and its robustness under numerous experimental variations.

In Eq.~\eqref{eqn:model1} $J_\textrm{loop}$ is the sum of the J-factors for each of four
possible loop configurations that have different DNA-binding
orientations, as well as for any additional loop conformations arising from protein flexibility (diagrammed in the legend of Fig. \ref{fig:Jtheory}).  The
J-factor depends on
  the length, phasing, and flexibility of the DNA, as well as the
  precise shape of the looped complex
  \cite{Towles2009,Swigon2006,Zhang2006}.  In fact, we observe two
looped states in almost all of our DNA constructs (see Fig.~\ref{fig:lengthtitr}(B) and (E)), as have other studies with Lac repressor
\cite{Han2009,Mehta1999,Edelman2003,Morgan2005,Wong2008,Normanno2008,Rutkauskas2009}.
Modifications to Eq.~\eqref{eqn:model1} that account for these multiple looped states,
as well as for experimental issues which may affect the $K_d$'s and J-factors we report, such as the
tetramer-to-dimer dissociation at low repressor concentrations, are
discussed in Section~\ref{sec:SImodels}.  However,
Eq.~\eqref{eqn:model1} is the main workhorse of the
paper since we found it to be sufficient to account
for the data presented here.  Similarly, in Sec.~\ref{sec:SIcontrols} we note a number of experimental controls that were performed to ensure that the parameters we fit to this model were not affected by the effects of the reporter bead size on loop formation, the large amount of surface area in the TPM sample chamber which
could cause a difference between the pipetted and actual
concentrations of repressor, or the particular repressor batch used in these experiments.

\subsection{Lac repressor purification.}  

As discussed in Sec.~\ref{sec:SIcontrols} in the Supplementary Material, we obtained reproducible TPM results only with Lac repressor purified in-house.  We used a protocol modified from one received from  Kathy Matthews in May 2009, essentially that described in \cite{Xu2009}.  The \emph{ E. coli lacI$^-$} BLIM cells and pJCI plasmid used for the purification were kind gifts from the Matthews lab.  After elution from the phosphocellulose column, our protein was found to have a concentration between 1 and 2 mg/mL, using a monomer extinction coefficient of 0.6 (mg/mL)$^{-1}$cm$^{-1}$ \cite{Butler1977}, and was $\ge$99\% pure by SDS-PAGE.  In one case some repressor was also purified over a Superdex 200 10/300 GL size-exclusion column (GE Healthcare) using an AKTA system and eluted as a single peak at a molecular weight corresponding to the expected weight of a LacI tetramer.

\subsection{DNAs.}

Plasmids pZS25' Oid-E/T(89-116)-O1$_{-45}$-YFP, where ``E/T(89-116)'' indicates that the sequence of the loop is either from the random E8 sequence or the 601TA sequence from \cite{Cloutier2005} and has a length of 89 to 116 bp, were constructed by site-directed mutagenesis as described in \cite{Han2009}.  Jonathan Widom kindly provided the E8 and TA sequences used in \cite{Cloutier2005}, which are a subset of those studied here and from which the other E8 and TA lengths were derived.  The operator and loop sequences used in this work can be found in Sec.~\ref{sec:DNAs}; schematics of the constructs without the {\it lac}UV5 promoter are shown in Fig.~\ref{fig:ExptSchematic}(B).  QuikChange site-directed mutagenesis (Agilent Technologies) was used to make the operator changes $O_{id}$ to $O_1$ and $O_{id}$ to $O_2$, additional loop lengths, and the promoter-containing constructs.  Linear labeled DNAs used in tethering assays were created by polymerase chain reaction (PCR) with primers labeled at the 5' ends with digoxigenin (forward primers) or biotin (reverse primers) (Eurofins MWG Operon); a PCR of the pZS25' plasmids resulted in approximately 450 bp tethers.  Primer sequences can be found in Table 3 of \cite{Han2009}.  See Fig.~\ref{fig:ExptSchematic}(B) for flanking DNA lengths for the no-promoter PCR products; the promoter-containing constructs of Fig.~\ref{fig:lengthtitr}(D-F) are identical to the no-promoter constructs shown in Fig.~\ref{fig:ExptSchematic}(B), except that the $O_1$ operator closest to the bead was replaced by $O_2$, 36 bp of the loop closest to this $O_2$ operator were replaced by the {\it lac}UV5 promoter sequence, and the length of the flanking DNA between $O_2$ and the bead was 139 bp rather than 172 bp.  

\subsection{TPM sample preparation, data acquisition and analysis.}

Our TPM protocol was essentially that of \cite{Han2009}, with the following modifications: \\
\\
\noindent(1) The addition of 0.2\% Tween-20 (Sigma) to the TPB buffer that some batches of beads were washed in, to reduce aggregation and nonspecific binding.  \\
\\
\noindent(2) Unless otherwise indicated, the beads used in this work were 0.49 $\mu$m-diameter, streptavidin-coated polystyrene beads (Bangs); 
for some controls in the Supplementary Material, 0.27 $\mu$m-diameter beads from Indicia Biotechnology 
were used instead.\\
\\
\noindent(3) Brightfield microscopy instead of differential interference contrast (the results are equivalent). \\
\\
\noindent(4) A Basler A602f camera was used to acquire images at a native frame rate of 60 frames per second (fps); however for consistency with previous results from our lab, every other frame was dropped for a final frame rate of 30 fps but an exposure time of 10 ms per frame.  \\
\\
\noindent(5) Improvements to the speed of the acquisition code that allowed up to 45 beads to be tracked at once, which corresponds to the maximal tether density obtainable in the field of view of the camera without a significant number of multiply tethered particles. \\
\\
\noindent(6) In addition to the symmetry-of-motion and length-of-motion checks that were used as initial screens for acceptable tethers in \cite{Han2009}, data were first acquired for 500 seconds in the Lac repressor buffer (LRB) but in the absence of protein in order to characterize each tether in the unlooped state.  Not only does this allow a more rigorous screening of tethers for anomalous behavior ({\it e.g.}, non-uniformity of tether length over time) but it also records the unlooped length of each individual bead, which allows easier identification of looped states, especially in DNAs with short loops that have high looping probabilities.  This must be done on a tether-by-tether basis due to the significant variability of  tether lengths that we see, and allows us to observe small differences in tether length in the presence versus absence of looping, which we attribute to operator bending (see Sec.~\ref{sec:DNAbending} in the Supplementary Material). \\
\\
\noindent(7) The non-covalent attachments of the DNAs to the surface and to the
bead can result in release of the tether from the surface before the conclusion of the experiment (usually about 1.5 hours in total).  As discussed in more detail in Sec.~\ref{sec:MinTimMinBds} in the Supplementary Material, beads that broke before 3000 seconds were excluded from the final analysis so
that all trajectories were sufficiently sampled to obtain the
equilibrium looping probability, and each data point includes at least 20
beads because fewer beads resulted in unreproducible looping
probabilities.  \\
\\
\noindent(8) In the case of the 0.49 $\mu$m beads, drift was removed
as described in \cite{Han2009} by subtracting the results of a
low-pass first-order Butterworth filter with a cutoff frequency 0.05
Hz; for the 0.27 $\mu$m beads, the cutoff frequency was 0.07 Hz.
Similarly, in the case of the 0.49 $\mu$m beads, the root-mean-square
motion was obtained by applying a Gaussian filter with a -3 dB
frequency of 0.0326 Hz, corresponding to a 4-second standard deviation
of the filter; but for the 0.27 $\mu$m beads, a 0.461 Hz filter was
used, corresponding to a 2.8-second standard deviation of the
filter. A 4-second Gaussian filter has a dead time of
  5.5 seconds; the temporal resolution of a TPM experiment is usually
  taken to be twice the dead time, or in our case, 11 seconds
  \cite{Finzi1995,Vanzi2006,Wong2008}.  The shortest-lived states that
  we observe have average lifetimes on the order of 30 seconds, which
  we so far have found to be long enough, compared to the temporal
  resolution imposed by the filter, as to make corrections for missed
  events negligible.  This issue will be addressed in more detail in a
  forthcoming paper on the kinetics of looping.\\
  \\
\noindent(9) We observe a population of tethers that never loop regardless of DNA construct or repressor concentration, and discarded these tethers from the calculation of the mean looping probability as described in Sec.~\ref{sec:NonLoopers} in the Supplementary Material.  \\
\\
\noindent(10) Fits were performed using custom Matlab routines as described in
Section~\ref{sec:SIfits}.  
Tracking and analysis code is available on request.  All data were obtained at 22-24$^o$C.  Looping probabilities are reported as means with standard errors; the calculation of looping J-factors and associated errors is described in Sec.~\ref{sec:dJexp} in the Supplementary Material.

\section{RESULTS}

\subsection{Effect of  repressor concentration and operator
strength on
looping probability.}\label{sec:operators}

We first explore how the Lac repressor concentration and its affinity for several known binding sites alter the looping probability, and how these alterations may be used to
extract the looping J-factor of the DNA, as well as the repressor-operator dissociation constants.    Looping by the Lac repressor has been studied by TPM
  \cite{Finzi1995,Vanzi2006,Wong2008,Normanno2008,Rutkauskas2009,Han2009,Milstein2011},
  as well as by other single-molecule techniques such as F\"orster resonance energy transfer (FRET)
  \cite{Mehta1999,Edelman2003,Morgan2005}, 
  but in all cases only one or a couple loop
  lengths, operators, and repressor concentrations were studied. In many cases therefore the
  repressor-operator dissociation constants were assumed (as opposed
  to measured) in order
  for a looping J-factor to be calculated.  Here we describe a new way of measuring both the operator dissociation constants and the relative
flexibilities of different DNA sequences as contained in the looping J-factor, by tuning both repressor concentration and operator strengths, with a rigorous comparison between these experiments and theory.  We find that the most accurate
and logically consistent way of measuring both the J-factors and operator dissociation constants involves a global fit of our model to multiple data sets with different combinations of operators simultaneously. 

\begin{figure}[htbp]
\begin{center}
\includegraphics[width=6.75in]{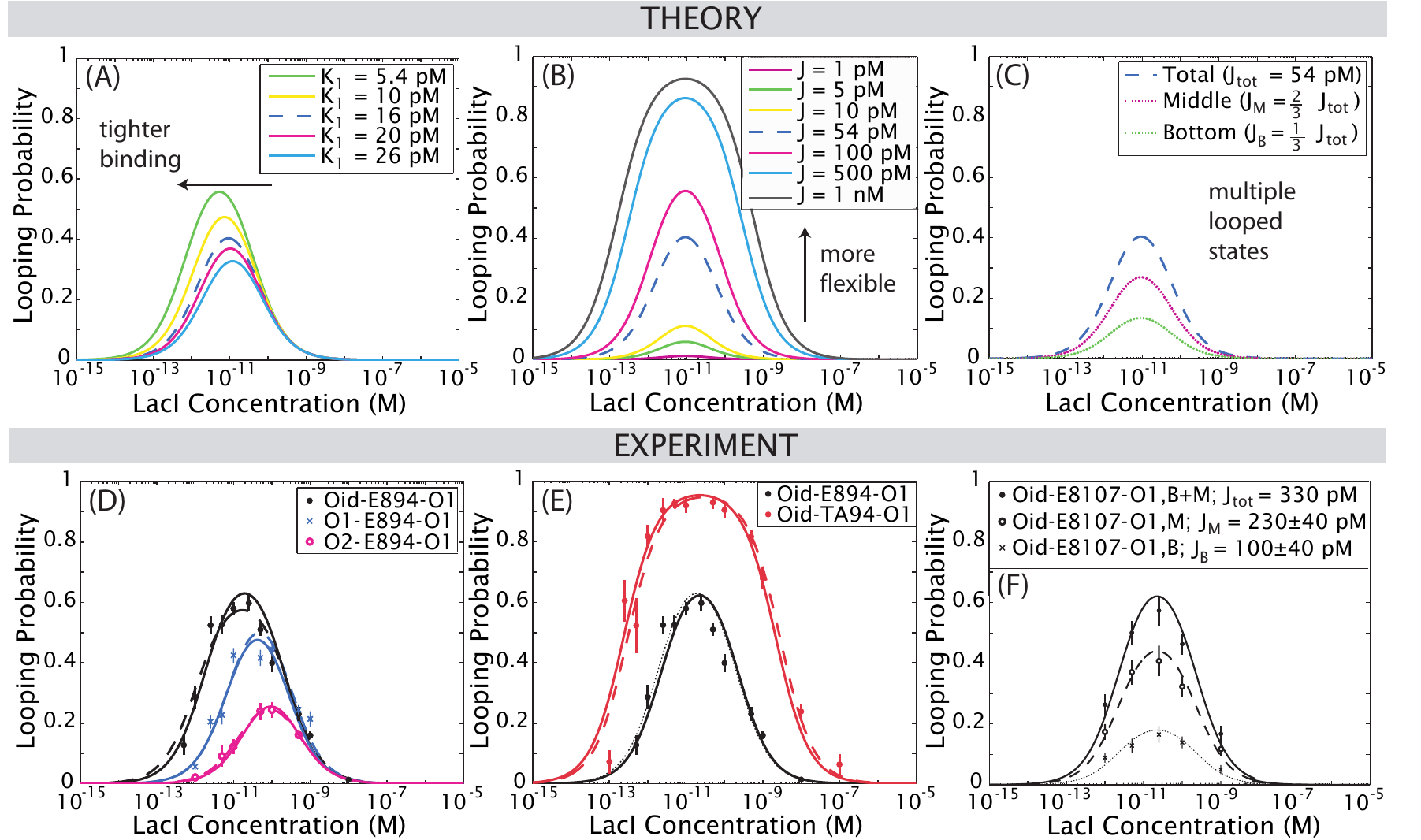}
\end{center}
\caption{Theoretical (A-C) and experimental (D-F) results for the looping probability as a function of operator strength, loop sequence, and repressor
  concentration.  In the theoretical predictions of (A)-(C), \protect{$\Kid$} = 5.4
  pM, $K_1$ = 16 pM, and \protect{$\J$} = 54 pM unless otherwise indicated; curves with these default parameters are shown as dashed blue lines for comparison across panels.  In the experimental results of (D)-(F),   unlike in (A-C), curves are fits to Eqn.~\eqref{eqn:model1}, not predictions.  
  {\bf (A)} Theoretical prediction for the effect
  of changing the strength of one of the operators on the looping probability as a function of repressor concentration. 
  {\bf (B)} Theoretical prediction for the effect of
  changing the flexibility of the DNA in the loop.  
  {\bf (C)}
  Extension of the simple model to the case of two experimentally
  distinguishable looped states (see Methods section and the section on loop length), which we model as having different
  J-factors.  The two looped states are labeled ``middle'' (``M'') and ``bottom'' (``B'') in reference to their relative tether lengths. 
 The dashed blue line shows the sum of the probabilities of the two
  states, which we refer to as the total looping probability.
  {\bf (D)} Measured looping probabilities for 94 bp of the random E8
  sequence, flanked by three different combinations of operators.
  Dashed lines indicate individual fits to each data set as described
  in Sec.~\ref{sec:SIfits} in the Supplementary Material; solid lines indicate a global fit to
  all three data sets simultaneously.  The global fit, which
    enforces identical values of the J-factor and $O_1$ dissociation constant
    in all three data sets, describes the data as well as the
  individual fits, demonstrating the consistency of the model when the
  operators are changed.  
  {\bf (E)} Looping probabilities for the E8 (black) and
    TA (red) sequences as function of concentration.  The Oid-E894-O1
    data are the same as in (D); the
     dotted black line is the result of the global fit
    shown in that panel as well.  The dashed red line represents an
    individual fit to the Oid-TA94-O1 data; the solid red and black
    lines are from a global fit to all three E8 data sets in
    (D) plus this TA data.  (The results of this global fit that includes the TA data for the O1-E894-O1 and O2-E894-O1 data sets are shown in Sec.~\ref{sec:SIfits} in the Supplementary Material.)  The TA data can be fit
    with the same $K_d$ values as the E8 data, but have a significantly
   larger J-factor, or a more flexible sequence.  Fit parameters for (D) and (E) are listed in
  Table~\ref{tab:fitparams}.
{\bf (F)} Looping probabilities for a DNA with two looped states, Oid-E8107-O1.  
Curves represent a simultaneous fit of the ``B'' and ``M'' data to Eqs.~\eqref{eqn:SItwostates1} and \eqref{eqn:SItwostates2}, using the values of \protect{$\Kid$} and $K_1$ from the global fit to all three E8 data sets in (D) and the TA data in (E). 
The procedure for determining the errors on the fit follows the bootstrapping scheme used throughout this work and is described in Sec.~\ref{sec:SIfits} in the Supplementary Material.  We find that the two looped states differ only in J-factor, as we and others \protect{\cite{Swigon2006,Zhang2006}} assume in our models; that is,  the binding affinity of the repressor for operator DNA does not change with the different loop and/or repressor conformations that generate the two observed loop states. 
Note that the total J-factor of 330 pM obtained from this concentration curve is within error of the J-factor of 280 $\pm$ 40 pM determined from only the 100 pM data point shown in Fig.~\ref{fig:lengthtitr}(C); 
likewise the J-factors for the two looped states are within error of those determined from the 100 pM data alone 
(Fig.~\ref{fig:SIJBvsM}(B)).}
\label{fig:E8TA}
\end{figure}

As described in the Materials and Methods section, we can use the tools of statistical mechanics to relate J-factors, operator dissociation constants, and transcription factor concentrations to the experimentally observable looping probability through the expression in Eq.~\eqref{eqn:model1}.
The main workhorse of our approach to test this statistical mechanical
description of looping probability is the repressor concentration curve, where we measure this probability at different repressor concentrations, and then fit Eq.~\eqref{eqn:model1} to obtain dissociation constants ($K_d$'s) and J-factors.  Equation~\eqref{eqn:model1} makes very specific and falsifiable predictions for how these repressor concentration curves should change as the model parameters change.   
Figure~\ref{fig:E8TA} shows a suite of previously untested predictions based upon this statistical mechanical model (as well as the comparison of these predictions to experiment).  We consider first the effect of changing the affinity of the repressor for its operators, and in the next section we consider the effect of changing the J-factor.
 
Figure~\ref{fig:E8TA}(A) shows the prediction of our model for how the concentration curves should change as the dissociation constant for one of the operators is varied: changing the
strength of one of the operators should change both the concentration at
which looping is maximal, and the amount of looping at that maximum,
but the curves should overlap at high repressor
concentrations.   These observations can be formalized
by appealing to Eq.~\eqref{eqn:model1}. 
 The concentration
at the maximum in the looping probability can be found by
differentiating Eq.~\eqref{eqn:model1} with respect to $[R]$ and results in
\begin{equation}
\Rmax = \sqrt{K_i K_{ii}}.\label{eqn:Rmax}
\end{equation}
Note that the concentration at which the
looping probability is maximized does not depend upon the DNA
flexibility as captured in the parameter $\J$.  The looping probability at this maximum, however, does depend on $\J$, according to
\begin{equation}
  \ploop(\Rmax) =
  \frac{\J/2}{\J/2+(\sqrt{K_i}+\sqrt{K_{ii}})^2},\label{eqn:ploopatRmax}
\end{equation}
and will therefore be discussed in more detail in the next section where our measurements of the J-factors of two different sequences are directly addressed.
Finally, we note that at high concentrations, Eq.~\eqref{eqn:model1} approaches the limit $\J/(2[R])$, which is independent of operator strength, explaining why the curves in Fig.~\ref{fig:E8TA}(A) overlap at high concentrations.  As an experimental consequence, data at low concentrations are essential for determining operator strengths, whereas high concentration data are sufficient for determining J-factors.

Figure~\ref{fig:E8TA}(D) shows experimental results for a loop containing 94 bp of a synthetic random sequence called E8, described previously \cite{Cloutier2004,Cloutier2005}, flanked by three different combinations of the operators $O_{id}$, $O_1$, and $O_2$, which are known to have distinct affinities for the Lac repressor.  
As predicted by our model, increasing the binding strength of one of the operators ({\it i.e.}, decreasing the value of one $K_d$) shifts the maximum of the curve to the left and increases its amplitude: that is, stronger operators allow more looping at lower concentrations. 
 Similarly, since the J-factor
is a property of the DNA loop length and sequence, we would expect all
three curves to be fit by the same J-factor, and for the fits to
reflect the reality that they share $O_1$ as one of the operators.  This
is indeed what we find, as shown in the fit parameters listed in
Table~\ref{tab:fitparams}: fits to the individual data sets (dashed
lines in Fig.~\ref{fig:E8TA}(D)) and a global fit to all three data
sets simultaneously (solid lines), where we have enforced the
constraint that all three data sets share the same J-factor and
dissociation constant of the $O_1$ operator, are
comparable in their fidelity.   We find that
the fitted values for the $K_d$'s agree well with values in the
literature obtained through bulk biochemical techniques (see
references cited in Table~\ref{tab:fitparams}), as well as for the
most part agreeing between individual fits to different data sets; and
that the fitted J-factor also agrees well between data sets, with a
value of about 300 $\pm$ 20 pM.  We are therefore confident that this combined concentration titration plus statistical mechanical model approach provides us with reasonable parameter values for both dissociation constants and J-factors, and that the global fit supplies the most reliable
  parameter estimates. 
  
 \begin{table}[tb]
\begin{tabular*}{\textwidth}{@{}c | c | c | c | c | c @{}}
\hline
Data & $K_{id}$  & $K_{1}$ & $K_2$ & $J_{\textrm{loop, E8}}$ & $J_{\textrm{loop, TA}}$ \\
 \hline
    Oid-E894-O1 & \phantom{1}3 ($\pm$ 1) & 90 ($\pm$ 20)           & --             & 350 ($\pm$ 40) & --\\
     O1-E894-O1 &                    --  & 47 ($\pm$ \phantom{0}4)  & --             & 380 ($\pm$ 30) & --\\
     O2-E894-O1 &                     -- & 26 (11, 125)            & 300 ($\pm$ 200)& 320 ($\pm$ 90) & --\\
    Oid-TA94-O1 &             10 (5, 46) & 80 ($\pm$ 40)           & --             & --             & 5500 ($\pm$ 600)\\
 Global Fit, E8 & \phantom{1}9 ($\pm$ 1) & 42 ($\pm$ \phantom{0}3) & 210 ($\pm$ \phantom{0}40) & 300 ($\pm$ 20) &--\\
 Global Fit, E8 \& TA &     12 ($\pm$ 3) & 44 ($\pm$ \phantom{0}3) & 240 ($\pm$ \phantom{0}50) & 330 ($\pm$ 30) & 4200 ($\pm$ 600)\\
\hline
   Literature values & 8.3$\pm$1.7 & 37$\pm$5 & 350$\pm$130 & -- & --\\
\hline
\end{tabular*}%
\caption{Measured dissociation
constants and looping J-factors, in pM, obtained by fitting
Eq.~\eqref{eqn:model1} to the data shown in Figs.~\ref{fig:E8TA}(D) and
(E).  In most cases the best fit parameter, plus or minus the standard
deviation of the distribution of fit parameters from bootstrapped
data, is reported; however in cases where the standard deviation
includes negative parameter values, a 95\% confidence interval is
reported in parentheses instead.  The first four rows are individual fits to the
indicated data sets; the fifth row is a global fit to all three of the
E8-containing data sets in Fig.~\ref{fig:E8TA}(D); and the sixth row
is a global fit to these three E8 data sets and the TA data set in Fig.~\ref{fig:E8TA}(E).
Fitting procedures are discussed in Sec.~\ref{sec:SIfits} in the Supplementary Material.  Literature values for \protect{$\Kid$} taken from Ref. \protect{\cite{Frank1997}}, for $K_1$ from Refs.  \protect{\cite{Hsieh1987,Whitson1986,WhitsonMatthews1986}}, and for $K_2$ from Ref. \protect{\cite{Hsieh1987}}.}
\label{tab:fitparams}
\end{table}

The looping J-factor for E894 is higher than the corresponding
cyclization J-factor of 54 pM reported in earlier work \cite{Cloutier2005}, and significantly higher than cyclization J-factors for other sequences of similar lengths \cite{Du2005}.  However, since the looped geometry imposes less stringent constraints
on the DNA than does cyclization (discussed in more detail below), we
would expect the looping J-factor to be larger than the cyclization
J-factor.   

\subsection{Effect of sequence on looping probability}\label{sec:E8vsTA}

Though the role of DNA sequence has not been extensively studied in the particular case of transcription-factor mediated looping, it has become a key parameter in the discussion of a different mechanism of transcriptional regulation, that of nucleosome positioning in eukaryotes \cite{Widom2001}. 
A number of sequences with very different nucleosome affinities have been identified, some isolated from natural sources and others from nucleosome affinity assays with synthetic sequences \cite{Widom2001}.  It has been argued for both classes that their nucleosomal affinities stem from different intrinsic flexibilities, and not in response to some other {\it in vivo}
condition or to a property specific to
  nucleosome binding, which in turn has led not only to many theoretical and experimental studies on the relationship between sequence and flexibility \cite{Peters2010,Hogan1983,Geggier2010,Olson2011}, but also to the determination of certain sequences that are claimed to be highly flexible.  
  For example, Cloutier and Widom characterized a sequence,
601TA, which has a significantly higher affinity for nucleosomes and a J-factor for cyclization 5 to 30 times
greater than the random E8 sequence described in the previous section, depending on the phasing discussed
in the next section \cite{Cloutier2004,Cloutier2005,Lowary1998}. 
 If 601TA and E8  differ in mechanical bendability in some general sense, then 601TA should increase looping by a bacterial transcription factor just
as it increases nucleosome binding and cyclizes more readily than E8.

As derived in Eqs.~\eqref{eqn:Rmax} and \eqref{eqn:ploopatRmax} and shown graphically in Fig.~\ref{fig:E8TA}(B), if the 601TA and E8 sequences have different
J-factors, then the concentration at which looping is maximal should
be the same for both sequences, but looping should increase at all
concentrations with the more flexible sequence.  This is indeed what
we find experimentally  in Fig.~\ref{fig:E8TA}(E), which shows results for the looping probability as a function of repressor concentration for a loop with 94 bp of a sequence derived
from 601TA (henceforth abbreviated to ``TA''), flanked by the $O_{id}$ and $O_1$ operators.
In analogy with the case of different
operators discussed in the previous section, the agreement between the
individual fit to the TA data (red dashed line) and the global fit to
both the E8 and TA data (solid lines) demonstrates that the two data
sets can be fit by the same operator dissociation
constants but different J-factors (see Table~\ref{tab:fitparams}).
The outcome of this measurement is a looping J-factor
  of 4.2 $\pm$ 0.6 nM for the TA sequence, about 10 times higher
  than the random E8 sequence.
This is again higher than the cyclization J-factors
  in~\cite{Cloutier2005} and \cite{Du2005} in terms of absolute magnitude, and significantly so: if we use Eq.~\eqref{eqn:ploopatRmax} and the cyclization J-factors of \cite{Cloutier2005} to predict maximal looping probabilities, we would expect  the maximal looping probability for Oid-E894-O1 to be 0.25 $\pm$ 0.3 (compared to the experimentally observed 0.62 $\pm$ 0.01), for Oid-TA94-O1 to be 0.87 $\pm$ 0.2 (compared to 0.95 $\pm$ 0.01), and the O2-E894-O1 construct to show essentially no looping at all.  The looping J-factor we measure for the TA sequence is not, however, as much higher than E8 as the 30-fold difference measured in
cyclization \cite{Cloutier2005}, hinting that the constraints imposed on the DNA in cyclization versus loop formation may lead to a different dependence on sequence, as indeed we find below. 

\subsection{Effect of loop length on looping probability}\label{sec:lengthtitr}

One of the signatures of looping by transcription factors both {\it in vitro} and {\it in
  vivo} is a significant modulation of transcription factor activity
as the distance between the transcription factor binding sites is
varied \cite{Schleif1992,Muller1996,Kramer1987,Bellomy1988}.  A similar phasing effect has
been observed in cyclization data with the E8 and TA sequences
\cite{Cloutier2005}.  
Our experiments, in conjunction with our model that allows us to extract J-factors, permit us to explore this
phasing behavior for both of the sequences discussed in the previous
section 
and to compare to several recent theoretical predictions of the looping J-factor.

\begin{figure}[p]
\begin{center}
\includegraphics[width=6.75in]{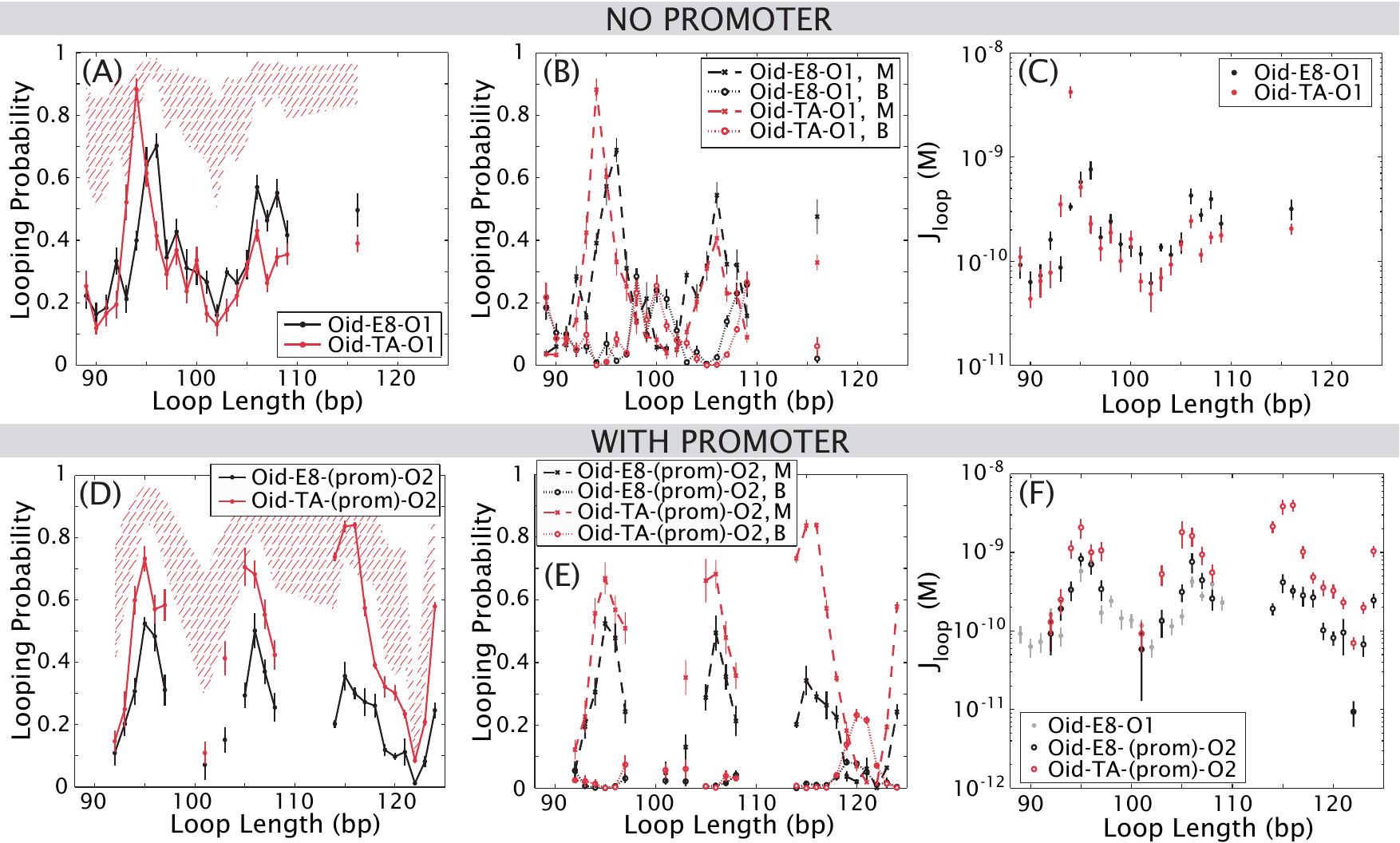}
\caption{Looping probability as a function of loop length at constant repressor concentration.  Surprisingly, the sequence dependence of Fig.~\ref{fig:E8TA}(E) for the 94 bp construct is absent at other loop lengths.  However, the bottom panels show data for constructs where 36 bp of either E8 or TA nearest O1 has been replaced with the {\it lac}UV5 promoter sequence (and for technical reasons O1 has been replaced with O2, which should not affect our measurements of J-factors as demonstrated by the data in Fig.~\ref{fig:E8TA}(D)).  The presence of this promoter {\it restores} a sequence dependence to looping across several helical periods.  
{\bf (A)} Total looping probability (that is, both looped states summed) for the constructs Oid-E8-O1 and Oid-TA-O1, at 100 pM repressor. The red hatched region represents a prediction for where the TA data should fall, assuming the TA sequence has a J-factor anywhere from 5 to 30 times larger than the J-factor for the E8 sequence (a range based on the cyclization J-factors of \protect{\cite{Cloutier2005}}).  The lengths used in earlier cyclization assays \protect{\cite{Cloutier2005}} are a subset of those shown in this figure. 
{\bf (B)} Looping probabilities for the two looped states separately (labeled ``bottom'' (``B'') and ``middle'' (``M'') as in Fig.~\ref{fig:E8TA}(C) and (F)) for the constructs in (A).  The two states alternate in likelihood: the bottom state predominates around 89 bp and 100 bp, but the middle state around 94 bp and 106 bp.  It is more clear in this panel than in (A) that E8 and TA are in phase with each other, with a period close to the canonical period of 10 bp, everywhere except near 94 bp, where TA has a maximum that is instead at 95-96 bp for E8.  Therefore a simple offset in phase between the two sequences cannot account for the behavior at 94 bp.
{\bf (C)} Looping J-factors for the constructs shown in (A).  The J-factors for both E8 and TA span at least an order of magnitude as a function of loop length, and the J-factors for the two looped states (see Fig.~\ref{fig:Jtheory} and Fig.~\ref{fig:SIJBvsM}(B) in the Supplementary Material) can also differ by an order of magnitude at a given loop length.  However, as shown in Fig.~\ref{fig:Jtheory}, this degree of modulation by operator phasing is less than might be predicted, depending on the assumptions made about Lac repressor conformation and flexibility.  
{\bf (D)} Looping probabilities for constructs where part of the looping sequence of the constructs in (A) has been replaced with the 36-bp {\it lac}UV5 promoter.  The red hatched region is the same kind of cyclization-based prediction as in (A). In sharp contrast to the data in (A), with the promoter sequence in the loop, TA loops as much or more than E8 at all lengths measured, as would be expected from cyclization and nucleosome formation assays with the pure E8 and TA sequences.  Note that because of the replacement of O1 by O2 the looping probabilities for these constructs will not necessarily match those of (A) even when the J-factors for the loops, plotted in (F), are the same.  
{\bf (E)} As in (B), here the two looped states have been separated out for the constructs in (D).  With the promoter in the loop, the two sequences have the same phasing even at 94 bp (and in fact share the same phasing as the pure E8 constructs in (A)).  Interestingly, the preferred looped state with the promoter is almost exclusively the middle state at all lengths---note for example that at 107 bp without the promoter, the two looped states are comparable in likelihood (see also Fig.~\ref{fig:E8TA}(F)), but with the promoter at 107 bp only the middle state contributes to looping (see also Fig.~\ref{fig:SIJBvsM}(D) and (E)).  
{\bf (F)} J-factors for the constructs in (D) (open circles), overlaid on the J-factors for
}
\label{fig:lengthtitr}
\end{center}
\end{figure}

In the spirit of the kinds of theoretical predictions of Fig.~\ref{fig:E8TA}(B), we can use the cyclization results of \cite{Cloutier2005}, which looked at the differences between E8 and TA across multiple DNA lengths, to make a na\"{i}ve prediction of how we would expect the sequence dependence to looping shown in Fig.~\ref{fig:E8TA}(E) to manifest as the loop length is changed.  Such a prediction is shown as a red hatched region in Fig.~\ref{fig:lengthtitr}(A). However, as shown in that figure, to our surprise our experimental results for the looping probabilities for the two sequences, at a constant
repressor concentration of 100 pM, show no sequence dependence to looping, with the exception of one or two lengths around the length shown in Fig.~\ref{fig:E8TA}(B).  The modulation of looping due to phasing is
observed in both the E8- and TA-containing sequences, and, with the exception of the 94 bp loop length, it appears that this phasing is the same for both sequences.  Yet again, surprisingly, not only does the nucleosome positioning sequence not fall within the hatched predicted region, in fact the nucleosome positioning sequence has comparable or smaller looping
probabilities compared to the random sequence at most loop lengths.

\newpage

\noindent{\footnotesize the no-promoter E8 construct shown in (C) (greyed-out closed circles).  The addition of the promoter to the loop does not appreciably change the J-factors for E8-containing loops, only those of the TA-containing loops. See Fig.~\ref{fig:SIJBvsM}(C) in the Supplementary Material for the J-factors of the two states of (E).  
Solid, dashed and dotted lines in (A), (B), (D), and (E) are guides to the eye only, not theoretical predictions or fits.  Their purpose is to highlight general trends.  Example bead motion-versus-time trajectories for these constructs can be found in
Section~\ref{sec:reptraces} in the Supplementary Material, and the effective tether lengths of the two looped states as a function of the loop length, with and without the promoter, are presented in Section~\ref{sec:DNAbending}.}

\noindent\hrulefill

Even more surprising is that a difference in loopability between the E8 and TA sequences can be restored when the last 36 bp of the loop is replaced with the bacterial {\it lac}UV5 promoter sequence, as shown in Fig.~\ref{fig:lengthtitr}(D).  We were motivated to make this change since in parallel work
we have measured how this sequence-dependent looping affects gene expression
{\it in vivo} and the presence of the promoter is a natural part of the
full regulatory network.   Though these loops contain 36 bp of the loop that are identical between the E8 and TA constructs, the TA-containing DNAs now loop more than the E8-containing DNAs, and at some lengths are even as much more flexible than the E8-containing DNAs as predicted based on cyclization assays, as shown by the red hatched region in Fig.~\ref{fig:lengthtitr}(D).  Interestingly, the J-factors for the E8 sequence with and without the promoter are comparable---that is, the inclusion of the promoter increases the flexibility of the TA-containing loops only (Fig.~\ref{fig:lengthtitr}(F)).

\begin{figure}[tbp]
\begin{center}
\includegraphics[width=3.45in]{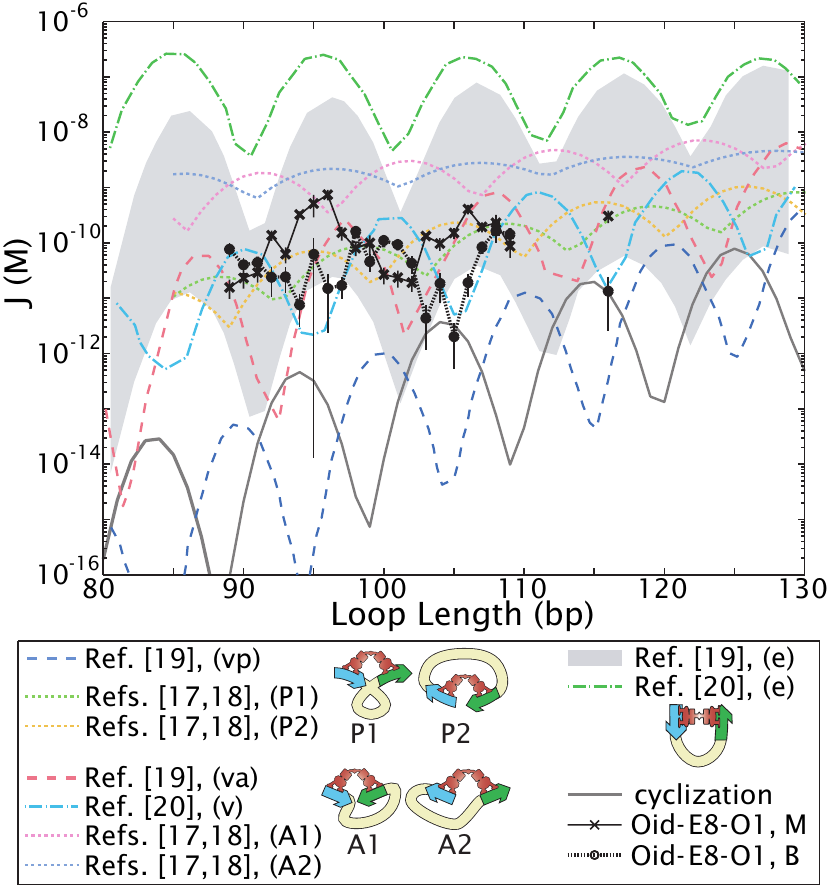} 
\end{center}
\caption{Theoretical predictions of the length dependence of the looping J-factor demonstrate that different assumptions about loop conformation and/or protein flexibility lead to predicted looping J-factors that can differ by orders of magnitude.  Elasticity theory with ``canonical'' values
    for the stiffness of random DNA sequences, in conjunction with
    various models of the geometric and mechanical constraints
    imposed by the Lac repressor tetramer, have been used to compute the looping
    J-factor \protect{\cite{Han2009,Towles2009,Swigon2006,Zhang2006}}. The model of \protect{\cite{Han2009,Towles2009}} also explicitly includes the boundary conditions of a TPM experiment, with a bead on one end of the DNA and a surface on the other.  The
    assumed constraints can be roughly grouped into V-like
    repressor conformations, similar to the shape seen in the crystal
    structure 1LBI \protect{\cite{Lewis1996}} (``P1'' and ``P2,'' indistinguishable unless as in TPM there are symmetry-breaking boundary conditions, and therefore collapsed into one state, ``vp,'' in \protect{\cite{Swigon2006}}; and ``A1'' and ``A2,'' collapsed into ``v'' or ``va'' in \protect{\cite{Swigon2006,Zhang2006}}); and more extended repressor conformations
    (``e''), which are favored by the DNA mechanics.  These conformations are indicated
    schematically in the legend; for the case of \protect{\cite{Han2009,Towles2009}}, the blue operator has been chosen to be $O_{id}$, that is, the operator closest to the surface.  The prediction for the extended conformation of \protect{\cite{Swigon2006}} is a range of values, reflecting estimated uncertainty in the free energy costs of opening the repressor tetramer.  
    Details of how these curves were obtained are given in Sec.~\ref{sec:Fig4Expl} in the Supplementary Material.  Our experimental
    measurements for the two looped states of the no-promoter E8 sequence (``Oid-E8-O1, M'' and ``Oid-E8-O1, B,'') as well as the cyclization result of Shimada and Yamakawa
    \protect{\cite{Shimada1984,Han2009}} (``cyclization'') have been included for
    comparison.  }
\label{fig:Jtheory}
\end{figure}

Before discussing the implications of these complex sequence dependencies,  we note several additional features of these length data in light of recent theoretical works on the length dependence of Lac repressor-mediated looping, which are plotted in Figure~\ref{fig:Jtheory}.  We and others observe two looped states with any pair of operators, which have been hypothesized to arise from the four distinct topological states of the looped DNA  and/or several distinct repressor conformations schematized in the legend of Fig.~\ref{fig:Jtheory} (see also the Methods section)
\cite{Wong2008,Normanno2008,Rutkauskas2009,Han2009,Mehta1999,Edelman2003,Morgan2005,Towles2009,Hirsh2011}. 
Regardless of their underlying molecular origins, in Fig.~\ref{fig:E8TA}(F) we show that the two looped states we observe can be modeled as differing only in effective J-factor; so in Fig.~\ref{fig:Jtheory} we compare the recent theoretical works plotted there to our experimental looping J-factors, but we do so for the two looped states separately, as each of the theoretical results make assumptions about the loop conformation that surely must differ between the two looped states we observe. As can be seen in that figure, different assumptions about the loop and protein geometry, and potential protein flexibility, lead to orders of magnitude differences in the predicted J-factors, reflecting our current uncertainty about the structure of the loop.  Moreover, no single theoretical work captures both the magnitude and the phasing of our experimental J-factors, suggesting that none of the theories accurately represents the loop structure yet.

We caution the reader, however, that a detailed direct comparison between these theoretical predictions and with our data 
    may not be possible for several reasons: (1) assumptions about experimental conditions such as salt concentrations differ between references and from the conditions in this work; (2) it is possible, as argued in \cite{Han2009,Towles2009}, that the experimentally observed states correspond to superpositions of two or more theoretically predicted states for different loop topologies and/or repressor conformations; and (3) as suggested by FRET data \cite{Edelman2003}, TPM with cross-linked repressor \cite{Rutkauskas2009}, and molecular dynamics simulations \cite{Villa2005}, the protein conformation in both states may involve some degree of rearrangement relative to the V-like conformation observed in the crystal structure (at the least, rotation of the DNA binding domains, as in \cite{Villa2005}).     In these cases our data would not align with any single theoretical curve.    However, we do make some general observations below and in Sec.~\ref{sec:Fig4Expl} in the Supplementary Material.

We find experimentally that the J-factors for the two states have opposite phasings, at least without the promoter, as shown in Fig.~\ref{fig:lengthtitr}(B), and this phasing does not change between sequences except near 94 bp. Such out-of-phase behavior for two different loop structures has been observed for other DNA looping proteins \cite{Watson2000}, and has been used to explain key features of {\it in vivo} repression data \cite{Saiz2007}. However it is not captured by all of the theoretical models in Fig.~\ref{fig:Jtheory} ({\it e.g.} the ``va'' and ``e'' states of Ref.~\cite{Swigon2006}).  Intriguingly, the promoter changes the relative probabilities of the two looped states: as shown in Fig.~\ref{fig:lengthtitr}(E), the promoter-containing constructs result almost exclusively in the middle state, whereas without the promoter, the two looped states alternate in prevalence (Fig.~\ref{fig:lengthtitr}(B)).  As these measurements represent the first single-molecule study on the phasing of these two looped states at single base-pair resolution, over two helical periods of DNA, at the short loop lengths where the models in Fig.~\ref{fig:Jtheory} show the most pronounced differences in J-factors due to repressor and loop conformations, we hope that our data will help shed light on the molecular origins of the two looped states. 

\section{DISCUSSION.}

We have shown here that the looping J-factors for 94 bp of a random sequence and a nucleosome positioning sequence differ by an order of magnitude, with the nucleosome positioning sequence being more flexible than the random sequence, as expected based on previous cyclization and nucleosome formation assays.  To our surprise, however, this sequence dependence occurs only at 94 bp, unless a bacterial promoter sequence is added to the loop, in which case a consistent length-independent sequence dependence is restored.

It is clear that data on more sequences are needed before any definitive conclusions can be drawn.  However we present here one possible hypothesis that we have considered: that the sequence-dependent free energy of bending a DNA depends more strongly than has been previously appreciated upon the specific details of how the DNA double helix is deformed when forming loops versus nucleosomes versus DNA circles. Drew and Travers argued that a DNA minicircle formed by cyclization shares structural similarities with the DNA wrapped around a histone octamer \cite{Drew1985}, explaining the usefulness of cyclization assays for understanding the sequence preferences of nucleosome formation.  Cyclization has often been cited as a model by which to understand looping as well \cite{Rippe1995,Cloutier2004,Geggier2010,Bellomy1990}.  However, as diagrammed in Fig.~\ref{fig:Jtheory}, for DNA loop formation by the Lac repressor, there are multiple looped configurations allowed for a given loop length, most of which are probably quite far from circular as a result of the distinct boundary conditions imposed by repressor binding, and which should have large effects on the associated looping J-factor.  We argue that although DNA cyclization may share characteristics with DNA looping such as length-dependent phasing, it apparently does not share other characteristics such as trends in sequence-dependent flexibility, possibly because of this difference in boundary conditions.

We also suspect that the strong sequence dependence at 94 bp without the promoter, and with the promoter at all lengths, is due to a change in the preferred loop conformations of these constructs, compared to the majority of the no-promoter constructs.  Indeed, the change in the predominant looped state (the ``bottom'' and ``middle'' states alternating without the promoter, but the ``middle'' state predominating at all lengths with the promoter) supports this hypothesis that the promoter alters the preferred conformation of the loop. Such a change in the preferred loop conformation could arise, for example, because of an intrinsic curvature to the {\it lac}UV5 promoter sequence.
To further unravel these subtleties we believe a high-throughput approach that
makes it possible to look at many sequences
might be necessary.  We also hope  that additional theoretical analyses, perhaps involving the observed tether lengths of the looped state with and without the promoter given in Section~\ref{sec:DNAbending} in the Supplementary Material, may shed further light on the conformations of looping for these different sequences. 

As discussed in the Introduction, the mechanics of loop formation at these short loop lengths that are so prevalent in cellular processes is a subject of much debate, regardless of their sequences \cite{Garcia2007,Peters2010}.   However, the question of how flexible we expect short DNAs to be is more complicated to answer in the case of protein-mediated DNA looping than in the case of cyclization.  As shown in Fig.~\ref{fig:Jtheory}, varying the boundary conditions of the loop or the assumed protein flexibility can lead to enormous differences in predicted looping J-factors.  Some of these predicted J-factors, using canonical assumptions about DNA flexibility, and without invoking anharmonic elasticity, are in fact consistent with the J-factors we measure, so perhaps it should not be surprising that short transcription factor-mediated loops can form readily {\it in vitro}.  

\section{CONCLUSION}

Transcription factor-mediated loops are a common motif in both prokaryotic and eukaryotic gene regulation.   Here, we have presented a combined single molecule plus modeling approach that allows us to explore how such looping is influenced by four distinct, tunable biological
parameters: transcription factor binding site strength, transcription factor concentration, DNA loop length,
and DNA loop sequence.  
We have demonstrated that this approach explains how the looping probability depends upon
the strength of the operator dissociation constants and that our measured 
$K_d$'s  agree well with values previously obtained by bulk biochemical methods.  Further, our model accounts well both quantitatively and qualitatively for the effects of varying 
the loop flexibility, as well as for details of our single-molecule looping experiments such as the presence of two looped states.  Our method provides a way of measuring
J-factors that is orthogonal to, and therefore complementary to, current methods in
  use, which we argue has led to important new insights into the role of sequence in DNA flexibility.  In particular we have argued here that the sequence-dependent free energy of bending a DNA must depend more strongly than has been previously appreciated upon the specific details of how the DNA double helix is deformed when forming loops versus nucleosomes versus DNA circles.  It is not the case that the TA sequence can be claimed to be more flexible in some general sense, nor can cyclization assays be used to determine DNA flexibility for all biological contexts, as we have shown here that loop formation does not necessarily follow the same sequence rules as cyclization.  Measurements of looping J-factors with many more sequences, and further theoretical explorations of the possible effects of sequence on these looping J-factors, will be necessary to understand the initial results presented here.
Continuing decades of work on the sequence-dependent mechanics of
DNA, the influence of sequence on DNA looping by transcription factors
now demands the same kind of scrutiny that has already been given to
nucleosome formation.

\section{SUPPLEMENTARY DATA}

Supplementary Data are available at NAR online: Supplementary tables 1-3, Supplementary figures 1-12, Supplementary methods, and Supplementary references \cite{Zhan2009,Barry1999,Chen1994,Chen1992,Brenowtiz1991,plischke2006,Oehler1994,newmanbarkema}.

\section{FUNDING}

This work was supported by the National Institutes of Health (NIH) [DP1 OD000217A (Director's Pioneer Award), R01 GM085286, R01 GM085286-01S1, and 1 U54 CA143869 (Northwestern PSOC Center)]; the National Science Foundation through a graduate fellowship (to S.J.); the Wenner-Gren foundation (to M.L.); the Fondation Pierre Gilles de Gennes (to R.P.); and the foundations of the Royal Swedish Academy of Sciences (to M.L.).  Funding for open access charge: NIH [1 U54 CA143869].

\section{ACKNOWLEDGEMENTS}

This work is dedicated to Jon Widom with warmth and appreciation for years of scientific
advice and insight including for this project.   We thank Kathy Matthews, Jia Xu, Kate Craig, Lin Han, Hernan Garcia,
 Phil Nelson, John Beausang, Laura Finzi, Jane Kondev, 
 Shimon Weiss, Bob Schleif, Dave Wu, Matthew
Johnson, Seth Blumberg, and Luke Breuer for insightful discussions and technical help,
and the Mayo, Shan, and Bjorkman labs for borrowed equipment and
advice on the LacI purification.

\subsection{Conflict of interest statement.}  None declared.



\begin{thebibliography}{10}

\bibitem{Echols1990}
Echols, H. (1990)
{Nucleoprotein structures initiating DNA replication, transcription, and
  site-specific recombination}.
{\em J Biol Chem,} {\bf 265}, 14697--14700.

\bibitem{Schleif1992}
Schleif, R. (1992)
{DNA looping}.
{\em Annu Rev Biochem,} {\bf 61}, 199--223.

\bibitem{Luijsterburg2006}
Luijsterburg, M.~S., Noom, M.~C., Wuite, G. J.~L., and Dame, R. T. (2006)
{The architectural role of nucleoid-associated proteins in the organization of
  bacterial chromatin: a molecular perspective}.
{\em J Struct Biol,} {\bf 156}, 262--272.

\bibitem{Matthews1992}
Matthews, K.~S. (1992)
{DNA looping}.
{\em Microbiol Rev,} {\bf 56}, 123--136.

\bibitem{Luger1997}
Luger, K., M\"{a}der, A.~W., Richmond, R.~K., Sargent, D.~F., and Richmond,
  T.~J. (1997)
{Crystal structure of the nucleosome core particle at 2.8~\AA~resolution}.
{\em Nature,} {\bf 389}, 251--260.

\bibitem{Garcia2007}
Garcia, H.~G., Grayson, P., Han, L., Inamdar, M., Kondev, J., Nelson, P.~C.,
  Phillips, R., Widom, J., and Wiggins, P.~A. (2007)
{Biological consequences of tightly bent DNA: the other life of a
  macromolecular celebrity}.
{\em Biopolymers,} {\bf 85}, 115--130.

\bibitem{Ptashne1986}
Ptashne, N. (1986)
{Gene regulation by proteins acting nearby and at a distance}.
{\em Nature,} {\bf 322}, 697--701.

\bibitem{Rippe1995}
Rippe, K., von Hippel, P.~H., and Langowski, J. (1995)
{Action at a distance: DNA-looping and initiation of transcription}.
{\em Trends Biochem Sci,} {\bf 20}, 500--506.

\bibitem{Tolhuis2002}
Tolhuis, B., Palstra, R.-J., Splinter, E., Grosveld, F., and de~Laat, W. (2002)
{Looping and interaction between hypersensitive sites in the active
  $\beta$-globin locus}.
{\em Molecular Cell,} {\bf 10}, 1453--1465.

\bibitem{Muller1996}
M\"{u}ller, J., Oehler, S., and M\"{u}ller-Hill, B. (1996)
{Repression of {\it lac} promoter as a function of distance, phase, and quality
  of an auxiliary {\it lac} operator}.
{\em J Mol Biol,} {\bf 257}, 21--29.

\bibitem{Peters2010}
Peters, J.~P. and Maher, L.~J. III. (2010)
{DNA curvature and flexibility {\it in vitro} and {\it in vivo}}.
{\em Quart Rev Biophys,} {\bf 43}, 22--63.

\bibitem{Olson2011}
Olson, W.~K. and Zhurkin, V.~B. (2011)
{Working the kinks out of nucleosomal DNA}.
{\em Curr Opin Struct Biol,} {\bf 21}, 348--357.

\bibitem{Schafer1991}
Schafer, D.~A., Gelles, J., Sheetz, M.~P., and Landick, R. (1991)
{Transcription by single molecules of RNA polymerase observed by light
  microscopy}.
{\em Nature,} {\bf 352}, 444--448.

\bibitem{Yin1994}
Yin, H., Landick, R., and Gelles, J. (1994)
{Tethered particle motion method for studying transcript elongation by a single
  RNA polymerase molecule}.
{\em Biophys J,} {\bf 67}, 2468--2478.

\bibitem{Jacobson1950}
Jacobson, H. and Stockmayer, W.~H. (1950)
{Intramolecular reaction in polycondensations. I. The theory of linear
  systems}.
{\em J Chem Phys,} {\bf 18}, 1600--1606.

\bibitem{Shimada1984}
Shimada, J. and Yamakawa, H. (1984)
{Ring-closure probabilities for twisted wormlike chains. Application to DNA}.
{\em Macromolecules,} {\bf 17}(4), 689--698.

\bibitem{Han2009}
Han, L., Garcia, H.~G., S., B., Towles, K.~B., Beausang, J.~F., Nelson, P.~C.,
  and Phillips, R. (2009)
{Concentration and length dependence of DNA looping in transcriptional
  regulation}.
{\em PLoS ONE,} {\bf 4}, e5621.

\bibitem{Towles2009}
Towles, K.~B., Beausang, J.~F., Garcia, H.~G., Phillips, R., and Nelson, P.~C.
  (2009)
{First-principles calculation of DNA looping in tethered particle experiments}.
{\em Phys Biol,} {\bf 6}, 025001.

\bibitem{Swigon2006}
Swigon, D., Coleman, B.~D., and Olson, W.~K. (2006)
{Modeling the Lac repressor-operator assembly: the influence of DNA looping on
  Lac repressor conformation}.
{\em Proc Natl Acad Sci USA,} {\bf 103}, 9879--9884.

\bibitem{Zhang2006}
Zhang, Y., {McEwen}, A.~E., Crothers, D.~M., Levene, S.~D., and Fraser, P.
  (2006)
Analysis of {\it in vivo} {LacR-mediated} gene repression based on the
  mechanics of {DNA} looping.
{\em PLoS ONE,} {\bf 1}, e136.

\bibitem{Mehta1999}
Mehta, R.~A. and Kahn, J.~D. (1999)
{Designed hyperstable Lac repressor$\cdot$DNA loop topologies suggest
  alternative loop geometries}.
{\em J Mol Biol,} {\bf 294}, 67--77.

\bibitem{Edelman2003}
Edelman, L.~M., Cheong, R., and Kahn, J.~D. (2003)
{Fluorescence resonance energy transfer over ~130 basepairs in hyperstable Lac
  repressor-DNA loops}.
{\em Biophys J,} {\bf 84}, 1131--1145.

\bibitem{Morgan2005}
Morgan, M.~A., Okamoto, K., Kahn, J.~D., and English, D.~S. (2005)
{Single-molecule spectroscopic determination of Lac repressor-DNA loop
  conformation}.
{\em Biophys J,} {\bf 89}(4), 2588--2596.

\bibitem{Wong2008}
Wong, O.~K., Guthold, M., Erie, D.~A., and Gelles, J. (2008)
{Interconvertible Lac repressor-DNA loops revealed by single-molecule
  experiments}.
{\em PLoS Biol,} {\bf 6}, e232.

\bibitem{Normanno2008}
Normanno, D., Vanzi, F., and Pavone, F.~S. (2008)
{Single-molecule manipulation reveals supercoiling-dependent modulation of {\it
  lac} repressor-mediated DNA looping}.
{\em Nucleic Acids Res,} {\bf 36}, 2505--2513.

\bibitem{Rutkauskas2009}
Rutkauskas, D., Zhan, H., Matthews, K.~S., Pavone, F.~S., and Vanzi, F. (2009)
{Tetramer opening in LacI-mediated DNA looping}.
{\em Proc Natl Acad Sci USA,} {\bf 106}, 16627--16632.

\bibitem{Xu2009}
Xu, J. and Matthews, K.~S. (2009)
{Flexibility in the Inducer Binding Region is Crucial for Allostery in the {\it
  Escherichia coli} Lactose repressor}.
{\em Biochemistry,} {\bf 48}, 4988--4998.

\bibitem{Butler1977}
Butler, A.~P., Revzin, A., and von Hippel, P.~H. (1977)
{Molecular parameters characterizing the interaction of {\it Escherichia coli
  lac} repressor with non-operator DNA and inducer}.
{\em Biochemistry,} {\bf 16}, 4757--4768.

\bibitem{Cloutier2005}
Cloutier, T.~E. and Widom, J. (2005)
{DNA twisting flexibility and the formation of sharply looped protein-DNA
  complexes}.
{\em Proc Natl Acad Sci USA,} {\bf 102}, 3645--3650.

\bibitem{Finzi1995}
Finzi, L. and Gelles, J. (1995)
{Measurement of Lactose repressor-mediated loop formation and breakdown in
  single DNA molecules}.
{\em Science,} {\bf 267}, 378--380.

\bibitem{Vanzi2006}
Vanzi, F., Broggio, C., Sacconi, L., and Pavone, F.~S. (2006)
{Lac repressor hinge flexibility and DNA looping: single molecule kinetics by
  tethered particle motion}.
{\em Nucleic Acids Res,} {\bf 34}, 3409--3420.

\bibitem{Milstein2011}
Milstein, J.~N., Chen, Y.~F., and Meiners, J.-C. (2011)
{Bead size effects on protein-mediated DNA looping in tethered-particle motion
  experiments}.
{\em Biopolymers,} {\bf 95}, 144--150.

\bibitem{Cloutier2004}
Cloutier, T.~E. and Widom, J. (2004)
{Spontaneous sharp bending of double-stranded DNA}.
{\em Mol Cell,} {\bf 14}, 355--362.

\bibitem{Du2005}
Du, Q., Smith, C., Shiffeldrim, N., Vologodskaia, M., and Vologodskii, A.
  (2005)
{Cyclization of short DNA fragments and bending fluctuations of the double
  helix}.
{\em Proc Natl Acad Sci USA,} {\bf 102}, 5397--5402.

\bibitem{Widom2001}
Widom, J. (2001)
{Role of DNA sequence in nucleosome stability and dynamics}.
{\em Quart Rev Biophys,} {\bf 34}, 1--56.

\bibitem{Hogan1983}
Hogan, M., LeGrange, J., and Austin, B. (1983)
{Dependence of DNA helix flexibility on base composition}.
{\em Nature,} {\bf 304}, 752--754.

\bibitem{Geggier2010}
Geggier, S. and Vologodskii, A. (2010)
{Sequence dependence of DNA bending rigidity}.
{\em Proc Natl Acad Sci USA,} {\bf 107}, 15421--15426.

\bibitem{Lowary1998}
Lowary, P.~T. and Widom, J. (1998)
{New DNA sequence rules for high affinity binding to histone octamer and
  sequence-directed nucleosome positioning}.
{\em J Mol Biol,} {\bf 276}, 19--42.

\bibitem{Kramer1987}
Kr\"{a}mer, H., Niem\"{o}ller, M., Amouyal, M., Revet, B., von
  Wilcken-Bergmann, B., and M\"{u}ller-Hill, B. (1997)
{{\it lac} repressor forms loops with linear DNA carrying two suitably spaced
  {\it lac} operators}.
{\em J Mol Biol,} {\bf 267}, 1305--1314.

\bibitem{Bellomy1988}
Bellomy, G.~R., Mossing, M.~C., and Record, M. T. Jr. (1988)
{Physical properties of DNA {\it in vivo} as probed by the length dependence of
  the {\it lac} operator looping process}.
{\em Biochemistry,} {\bf 27}, 3900--3906.

\bibitem{Hirsh2011}
Hirsh, A.~D., Lillian, T.~D., Lionberger, T.~A., and Perkins, N.~C. (2011)
{DNA modeling reveals an extended Lac repressor conformation in classic {\it in
  vitro} binding assays}.
{\em Biophys J,} {\bf 101}, 718--726.

\bibitem{Villa2005}
Villa, E., Balaeff, A., and Schulten, K. 
{Structural dynamics of the lac {repressor{\textendash}DNA} complex revealed by
  a multiscale simulation}.
{\em Proc Natl Acad Sci USA,} {\bf 102}, 6783 --6788.

\bibitem{Watson2000}
Watson, M.~A., Gowers, D.~M., and Halford, S.~E. (2000)
{Alternative geometries of DNA looping: an analysis using the {\it Sfi}I
  endonuclease}.
{\em J Mol Biol,} {\bf 298}, 461--475.

\bibitem{Saiz2007}
Saiz, L. and Vilar, J. M.~G. (2007)
{Multilevel deconstruction of the {\it in vivo} behavior of looped DNA-protein
  complexes}.
{\em PLoS ONE,} {\bf 2}, e355.

\bibitem{Drew1985}
Drew, H.~R. and Travers, A.~A. (1985)
{DNA bending and its relation to nucleosome positioning}.
{\em J Mol Biol,} {\bf 186}, 773--790.

\bibitem{Bellomy1990}
Bellomy, G.~R. and Record, M. T. Jr. (1990)
{Stable DNA loops {\it in vivo} and {\it in vitro}: roles in gene regulation at
  a distance and in biophysical characterization of DNA}.
{\em Prog Nucleic Acid Res Mol Biol,} {\bf 39}, 81--128.

\bibitem{Lewis1996}
Lewis, M., Chang, G., Horton, N.~C., Kercher, M.~A., Pace, H.~C., Schumacher,
  M.~A., Brennan, R.~G., and Lu, P. (1996)
Crystal structure of the Lactose operon repressor and its complexes with {DNA}
  and inducer.
{\em Science,} {\bf 271}, 1247--1254.

\bibitem{Frank1997}
Frank, D.~E., Saecker, R.~M., Bond, J.~P., Capp, M.~W., Tsodikov, O.~V.,
  Melcher, S.~E., Levandoski, M.~M., and Record, M.~T. Jr. (1997)
{Thermodynamics of the interactions of {\it lac} repressor with variants of the
  symmetric {\it lac} operator: Effects of converting a consensus site to a
  non-specific site}.
{\em J Mol Biol,} {\bf 267}, 1305--1314.

\bibitem{Hsieh1987}
Hsieh, W.-T., Whitson, P.~A., Matthews, K.~S., and Wells, R.~D. (1987)
{Influence of sequence and distance between two operators on interaction with
  the {\it lac} repressor}.
{\em J Biol Chem,} {\bf 262}, 14583--14591.

\bibitem{Whitson1986}
Whitson, P.~A., Olson, J.~S., and Matthews, K.~S. (1986)
{Thermodynamic analysis of the Lactose repressor-operator DNA interaction}.
{\em Biochemistry,} {\bf 25}, 3852--3858.

\bibitem{WhitsonMatthews1986}
Whitson, P.~A. and Matthews, K.~S. (1986)
{Dissociation of the Lactose repressor-operator DNA complex: Effects of size
  and sequence context of operator-containing DNA}.
{\em Biochemistry,} {\bf 25}, 3845--3852.

\bibitem{Zhan2009}
Zhan, H., Sun, Z., and Matthews, K.~S. (2009)
{Functional impact of polar and acidic substitutions in the Lactose repressor
  hydrophobic monomer-monomer interface with a buried lysine.}.
{\em Biochemistry,} {\bf 48}(6), 1305--1314.

\bibitem{Barry1999}
Barry, J.~K. and Matthews, K.~S. (1999)
{Thermodynamic analysis of unfolding and dissociation in Lactose repressor
  protein}.
{\em Biochemistry,} {\bf 38}, 6520--6528.

\bibitem{Chen1994}
Chen, J. and Matthews, K.~S. (1994)
{Subunit dissociation affects DNA binding in a dimeric {\it Lac} repressor
  produced by C-terminal deletion.}.
{\em Biochemistry,} {\bf 33}, 8728--8735.

\bibitem{Chen1992}
Chen, J. and Matthews, K.~S. (1992)
{Deletion of Lactose repressor carboxyl-terminal domain affects tetramer
  dissociation.}.
{\em J Biol Chem,} {\bf 267}(20), 13843--13850.

\bibitem{Brenowtiz1991}
Brenowitz, M., Mandal, N., Pickar, A., Jamison, E., and Adhya, S. (1991)
{DNA-binding properties of a Lac repressor mutant incapable of forming
  tetramers.}.
{\em J Biol Chem,} {\bf 266}(2), 1281--1288.

\bibitem{plischke2006}
Plischke, M. and Birger, B. (2006)
Equilibrium statistical physics,
World Scientific, Hackensack {NJ} 3rd edition.

\bibitem{Oehler1994}
Oehler, S., Amouyal, M., Kolkhof, P., von Wilcken-Bergmann, B., and
  M\"{u}ller-Hill, B. (1994)
{Quality and position of the three {\it lac} operators of {\it E. coli} define
  efficiency of repression.}.
{\em EMBO J,} {\bf 13}(14), 3348--3355.

\bibitem{newmanbarkema}
Newman, M. E.~J. and Barkema, G.~T. (1999)
Monte Carlo Methods in Statistical Physics,
Oxford University Press, {USA}.

\bibitem{Segall2006}
Segall, D.~E., Nelson, P.~C., and Phillips, R. (2006)
{Volume-exclusion effects in tethered-particle experiments: bead size
  matters.}.
{\em Phys Rev Lett,} {\bf 96}(8), 088306.

\end{thebibliography}
\end{document}